\documentclass[prd,twocolumn,showpacs,superscriptaddress]{revtex4}

\usepackage{amsfonts}
\usepackage{amsmath}
\usepackage{amssymb}
\usepackage{amsthm}
\usepackage{bm}
\usepackage{dcolumn}
\usepackage{epsfig}
\usepackage{graphicx}
\usepackage{graphics}
\usepackage[latin1]{inputenc}
\usepackage{latexsym}
\usepackage{rotating}
\usepackage{hyperref}

\newcommand{\beqs}{\begin{equation*}}
\newcommand{\beq}{\begin{equation}}

\newcommand{\eeqs}{\end{equation*}}
\newcommand{\eeq}{\end{equation}}

\newcommand{\beqas}{\begin{eqnarray*}}
\newcommand{\beqa}{\begin{eqnarray}}

\newcommand{\eeqas}{\end{eqnarray*}}
\newcommand{\eeqa}{\end{eqnarray}}

\newcommand{\eq}[2]{\begin{equation} #1 \label{#2} \end{equation}}

\newcommand{\eps}{\varepsilon}
\newcommand{\al}{\alpha}
\newcommand{\be}{\beta}
\newcommand{\ga}{\gamma}
\newcommand{\de}{\delta}
\newcommand{\om}{\omega}

\newcommand{\la}{\lambda}
\newcommand{\si}{\sigma}

\newcommand{\Ga}{\Gamma}

\newcommand{\Om}{\Omega}

\newcommand{\blist}{\begin{itemize}}

\newcommand{\elist}{\end{itemize}}

\providecommand{\href}[2]{#2}

\DeclareFontFamily{OT1}{rsfs}{}
\DeclareFontShape{OT1}{rsfs}{m}{n}{ <-7> rsfs5 <7-10> rsfs7 <10->rsfs10}{} 
\DeclareMathAlphabet{\mycal}{OT1}{rsfs}{m}{n}

\DeclareMathOperator{\extdm}{d}
\newcommand{\extd}{\extdm \!}

\newcommand{\exponent}{b} 
\newcommand{\PMR}{J} 

\newcommand{\aff}{{\cal J}}
\newcommand{\vir}{{\cal L}}

\begin{document}
\title{Holograms of Conformal Chern--Simons Gravity}

\author{Hamid Afshar}
\email{afshar@ipm.ir}
\affiliation{Institute for Theoretical Physics, Vienna University of Technology, Wiedner Hauptstr. 8-10/136, A-1040 Vienna, Austria, Europe}
\affiliation{School of Physics, Institute for Research in Fundamental Sciences (IPM), P.~O. Box 19395-5531, Tehran, Iran}
\affiliation{Department of Physics, Sharif University of Technology, P.~O. Box 11365-9161, Tehran, Iran}

\author{Branislav Cvetkovi{\'c}}
\email{cbranislav@ipb.ac.rs}
\affiliation{University of Belgrade, Institute of Physics, P.~O. Box 57, 11001 Belgrade, Serbia}

\author{Sabine Ertl}
\email{sertl@hep.itp.tuwien.ac.at}
\affiliation{Institute for Theoretical Physics, Vienna University of Technology, Wiedner Hauptstr. 8-10/136, A-1040 Vienna, Austria, Europe}

\author{Daniel Grumiller}
\email{grumil@hep.itp.tuwien.ac.at}
\affiliation{Institute for Theoretical Physics, Vienna University of Technology, Wiedner Hauptstr. 8-10/136, A-1040 Vienna, Austria, Europe}

\author{Niklas Johansson}
\email{niklasj@hep.itp.tuwien.ac.at}
\affiliation{Institute for Theoretical Physics, Vienna University of Technology, Wiedner Hauptstr. 8-10/136, A-1040 Vienna, Austria, Europe}

\date{\today}

\preprint{TUW-11-xx}

\begin{abstract}
We show that conformal Chern--Simons gravity in three dimensions has various holographic descriptions. 
They depend on the boundary conditions on the conformal equivalence class and the Weyl factor, even when the former is restricted to asymptotic Anti-deSitter behavior.
For constant or fixed Weyl factor our results agree with a suitable scaling limit of topologically massive gravity results.
For varying Weyl factor we find an enhancement of the asymptotic symmetry group, the details of which depend on certain choices.
We focus on a particular example where an affine $\hat u(1)$ algebra related to holomorphic Weyl rescalings shifts one of the central charges by 1.  
The Weyl factor then behaves as a free chiral boson in the dual conformal field theory.
\end{abstract}


\maketitle


Conformal Chern--Simons gravity (CSG) \cite{Deser:1982vy,
Horne:1988jf} is a 3-dimensional third-derivative theory of gravity that has a parity-odd non-covariant action
\eq{
S_{\textrm{CSG}}=\frac{k}{4\pi}\,\int\!\extd^3x\, \epsilon^{\la\mu\nu}\,\Ga^\si_{\la\rho}\,\Big(\partial_\mu\Ga^\rho_{\nu\si}+\tfrac23\,\Ga^\rho_{\mu\tau}\Ga^\tau_{\nu\si}\Big)
}{eq:gCS10}
but covariant equations of motion
\eq{
C_{\mu\nu}:=\eps_\mu{}^{\la\si}\nabla_\la\big(R_{\nu\si}-\frac14\,g_{\nu\si}R\big)=0\,.
}{eq:gCS1}
Here $k$ is the Chern--Simons level, which we assume to be some positive integer, $\Ga$ are the Christoffel symbols and $C_{\mu\nu}$ is the Cotton tensor, the vanishing of which is equivalent to conformal flatness in three dimensions.
In the bulk the theory \eqref{eq:gCS10} is not only diffeomorphism invariant but also invariant under local Weyl rescalings
\eq{
g_{\mu\nu}\to e^{2\Om} g_{\mu\nu}\,.
}{eq:Weyl}
Consequently, the theory defined by the action \eqref{eq:gCS10} is topological in the sense that it has zero physical bulk degrees of freedom.
Interesting physical properties emerge if a boundary
is introduced \cite{Brown:1986nw,Witten:1988hc} --- for instance an asymptotic boundary, like in the holographic AdS/CFT correspondence \cite{Maldacena:1997re,Aharony:1999ti}. 

In this paper we show that CSG allows for various qualitatively different holographic conformal field theory (CFT) duals, depending on the boundary conditions imposed on the Weyl factor.
We focus on presenting and discussing the main results, and we shall provide a more detailed account of the calculations elsewhere \cite{ABEGJ:2011}.


We assume that the manifold $M$ has a (connected) boundary $\partial M$ with cylindrical or toric topology; $\partial M$ may but need not be an asymptotic boundary.
It is convenient to parametrize the boundary such that one of the coordinates, $y$, is constant on it.
With no loss of generality we assume $y=0$ at $\partial M$.
In the vicinity of the boundary we write the metric $g_{\mu\nu}$ as
\eq{
g_{\mu\nu} = e^{2\phi}\,\bar{g}_{\mu\nu} = e^{2\phi} \big( g_{\mu\nu}^{\rm AAdS} + h_{\mu\nu}\big)
}{eq:bc1}
and impose the condition that the metric $\bar{g}_{\mu\nu}$ be asymptotically AdS. More specifically,
with the leading metric 
\eq{
g_{\mu\nu}^{\rm AAdS}\,\extd x^\mu\extd x^\nu = \frac{\extd x^+\extd x^- + \extd y^2}{y^2}
}{eq:bc2}
we require that the subleading state-dependent part $h_{\mu\nu}$ take the form
\eq{
\left(\begin{array}{lll}
h_{++}={\cal O}(1/y) & h_{+-} = {\cal O}(1) & h_{+y} = {\cal O}(1) \\
              & h_{--} = {\cal O}(1)        & h_{-y} = {\cal O}(1) \\
              &                             & h_{yy} = {\cal O}(1)                      
\end{array}\right)\,.
}{eq:bc3}
The boundary conditions \eqref{eq:bc2} with \eqref{eq:bc3} restrict the conformal equivalence class of the metric.
The Weyl factor $\phi$ has to be considered separately.
Depending on its properties we distinguish three cases:
\begin{enumerate}
 \item[I.] Trivial Weyl factor $\phi=\rm const.$ ($=0$)
 \item[II.] Fixed Weyl factor $\phi\neq \rm const.$
 \item[III.] Free Weyl factor $\phi$ not fixed completely
\end{enumerate}
The reason for our split into boundary conditions on the conformal class and on the Weyl factor is the enhanced gauge invariance of CSG:
if $g$ is a solution to the equations of motion \eqref{eq:gCS1} then also $e^{2\phi}\,g$ is a solution.
The boundary conditions on the conformal class of the metric, \eqref{eq:bc2}-\eqref{eq:bc3}, are chosen such that AdS is allowed as a background and that most of the linearized excitations around AdS are admissible.
However, we cannot consistently allow all such excitations.
The rationale behind the precise choices above will be explained elsewhere \cite{ABEGJ:2011}.
For the present context it is sufficient to point out that the boundary conditions \eqref{eq:bc1}-\eqref{eq:bc3} should not be made stronger, since this would eliminate interesting solutions, and cannot be made looser, since this would lead to inconsistencies, like infinite charges.

\section{Trivial Weyl factor}

In this section we focus on case~I.
Since all the results in this section turn out to coincide with suitable scaling limits of topologically massive gravity (TMG) the presentation will be condensed, and we refer to \cite{ABEGJ:2011} for a more detailed analysis.
The boundary conditions \eqref{eq:bc3} are preserved by diffeomorphisms generated by a vector field $\xi$ with the properties
\begin{subequations}
 \label{eq:gCS28}
\begin{align}
 \xi^\pm &= \eps^\pm(x^\pm) - \frac12\,y^2\,\partial_\mp^2\eps^\mp(x^\mp) + {\cal O}(y^3)\,,\\
 \xi^y &= \frac y2\,\big(\partial_+\eps^+ + \partial_-\eps^-\big) + {\cal O}(y^3)\,.
\end{align}
\end{subequations}
They also allow asymptotic Weyl rescalings \eqref{eq:Weyl} with
\eq{
\Om = {\cal O}(y^2)\,.
}{eq:gCS31}
These Weyl rescalings are trivial symmetries, which are modded out in the asymptotic symmetry algebra.

We calculate now the response functions using the standard AdS/CFT dictionary \cite{Aharony:1999ti}.
To this end we need the first variation of the on-shell action $\de S=\de S_{\rm CSG}|_{\rm EOM}+\de S_b|_{\rm EOM}$, including appropriate boundary terms $S_b$, which we take from \cite{Guica:2010sw}. The result is
\eq{
\de S = \frac{1}{2}\,\int_{\partial M}\!\!\!\extd^2x\sqrt{-\ga^{(0)}}\,\Big(T^{\al\be}\,\de\ga^{(0)}_{\al\be} + \PMR^{\al\be}\,\de\ga^{(1)}_{\al\be}\Big).
}{eq:gCS45}
For convenience we use Gaussian normal coordinates in the asymptotic expansion ($e^\rho\propto 1/y$)
\eq{
\extd s^2 = \extd \rho^2 + \big(\ga^{(0)}_{\al\be}\,e^{2\rho}+\ga^{(1)}_{\al\be}\,e^\rho+\ga^{(2)}_{\al\be}+\dots\big)\,\extd x^\al\extd x^\be\,,
}{eq:FG}
where $\ga^{(0)}$ is the boundary metric, $\ga^{(1)}$ describes (partially massless) Weyl gravitons and their sources, and $\ga^{(2)}$ contains information about the left- and right-moving massless boundary gravitons.
The appearance of $\ga^{(1)}$ is the only difference to the situation studied by Brown and Henneaux in their seminal paper \cite{Brown:1986nw}.
The response functions $T^{\al\be}$ and $\PMR^{\al\be}$ are Brown--York stress tensor and partially massless response, respectively.
They correspond to operators in the dual CFT and are given by \cite{ABEGJ:2011}
\begin{align}
\label{eq:T}
 T^{\al\be} &= \frac{k}{\pi}\,\eps^{(\al}{}_\ga\,\ga^{\be)\ga}_{(2)} = \frac{k}{2\pi}\,\eps^{\al\ga}\,\ga^{\be\,(2)}_{\;\,\ga}+ (\al\leftrightarrow\be)\,, \\ 
 \PMR^{\al\be} &= \frac{k}{4\pi}\,\big(\de^{(\al}_{\ga}-\eps^{(\al}{}_{\ga}\big)\,\big(\ga^{\be)\ga}_{(1)}-\frac12\,\ga^{\be)\ga}_{(0)}\ga_{\si\de}^{(1)}\ga^{\si\de}_{(0)}\big) \,. 
\label{eq:J}
\end{align}
The results \eqref{eq:T}, \eqref{eq:J} for the 1-point functions agree with corresponding results in TMG \cite{Kraus:2005zm,Skenderis:2009nt}.
The same applies to 2-point functions \cite{Skenderis:2009nt} and 3-point functions \cite{Grumiller:2009mw}.
The non-vanishing 2-point functions are given by [$T^{R/L}$ are the (anti-)holomorphic flux components of the stress-energy tensor, with $x^{\pm}=\varphi\pm t$, $z=\varphi+it$]:
\begin{align}
& \langle\PMR(z,\bar z)\PMR(0,0)\rangle = \frac{2k\,\bar z}{z^3} \label{eq:JJ} \\
& \langle T^R(z)T^R(0)\rangle = \frac{6k}{z^4} = -\overline{\langle T^L(\bar z)T^L(0)\rangle} \label{eq:TT}
\end{align}
The result \eqref{eq:JJ} shows that one of the conformal weights of the partially massless Weyl gravitons is negative, $\bar h=-1/2$, in agreement with the classical \cite{Grumiller:2010tj} and 1-loop analysis \cite{Bertin:2011jk}.
From the 2-point functions \eqref{eq:TT} we can read off the central charges of the dual CFT.
\eq{
c_{R/L} = \pm 12 k 
}{eq:gCS54}
The result \eqref{eq:gCS54} agrees with a suitable limit of the TMG results for the central charges \cite{Kraus:2005zm,Solodukhin:2005ah}.

\section{Fixed Weyl factor}

Let us now address case~II where the Weyl factor $\phi$ in \eqref{eq:bc1} is arbitrary but fixed.
In order to recover the desired asymptotic diffeomorphisms \eqref{eq:gCS28} it turns out that we need to restrict its asymptotic behavior to~\footnote{%
It is possible to allow also a logarithmically divergent term $\exponent\,\ln{y}$ but with some constant $\exponent$ \cite{ABEGJ:2011}. 
We omit its discussion in the present work.
Allowing for such a term could lead to new anomalous contributions in the symmetry algebra in case~III.}
\eq{
\phi(x^+,\,x^-,\,y) = f(x^+,\,x^-) + \ldots
}{eq:gCS30}
We then find that the asymptotic diffeomorphisms \eqref{eq:gCS28} have to be accompanied by a compensating infinitesimal Weyl rescaling \eqref{eq:Weyl} with
\eq{
\Om = - \big(\eps^+\partial_+ + \eps^-\partial_-\big)\,f + {\cal O}(y^2)\,.
}{eq:gCS29}
Thus, the boundary condition preserving gauge transformations are precisely as for case~I, but we have to simultaneously rescale the metric with a Weyl factor \eqref{eq:gCS29}.

The only essential difference to case~I is the result for the Brown--York stress tensor, which no longer is conserved.
\eq{
\nabla_{\al} T^{\al\be} 
\propto \eps^{\be\ga}\partial^\al\partial_\al\partial_\ga f
}{eq:an7}
This effect was explained in \cite{Kraus:2005zm}.
On the gravity side, the reason for the anomalous conservation \eqref{eq:an7} is that the action is not diffeomorphism invariant: it transforms
by a boundary term. This anomaly vanishes only for flat boundary metrics. 
Indeed, requiring the stress tensor to be conserved implies [we drop a term proportional to $x^+x^-$ since it is not periodic in $\varphi = (x^+ + x^-)/2$]. 
\eq{
f(x^+, x^-) = f_+(x^+) + f_-(x^-)\,.
}{eq:an7.5}
Restricting to this class of Weyl factors $\phi$ produces a class of diffeomorphism anomaly free CFTs.

Interestingly, and not unexpectedly, the potential non-conservation of the charges implied by the Brown--York analysis above is invisible in a canonical analysis.
The difference between the Brown--York and the canonical result comes about because the former is based upon the variation of the action \eqref{eq:gCS10}, 
which is not diffeomorphism invariant at the boundary, while the latter is based upon the first order action ($T_i:=de_i+\eps_{ijk}\om^j e^k$)
\eq{
S^{(1)}=\frac{k}{2\pi}\,\int\Big[\om^i\extd\om_i +\tfrac{1}{3}\eps_{ijk}\,\om^i\om^j\om^k + \la^i T_i\Big] 
}{2.1}
which is manifestly diffeomorphism invariant. See again \cite{Kraus:2005zm}.
A key result of the canonical analysis \cite{ABEGJ:2011} (using the same methods as in \cite{Blagojevic:2010ir}) is an expression for the diffeomorphism charges $Q^P[\xi^\rho]$ 
($\de Q$ denotes the difference in charge between two states in the theory):
\begin{multline}
\de Q_P[\xi^\rho] = -\frac{k}{2\pi}\,\int\limits_0^{2\pi} \extd \varphi \, \Big[ \xi^\rho \big(e^i{}_\rho\, \de \la_{i\varphi}  + \la^i{}_\rho\, \de e_{i\varphi}  \\
+ 2 \om^i{}_\rho\, \de \om_{i\varphi} \big) +  2\theta^i[\xi^\rho]\, \de\om_{i\varphi} \Big]\, . \label{eq:diffcharge}
\end{multline}
The last term proportional to the Lorentz parameter $\theta^i[\xi^\rho]$ vanishes asymptotically for cases~I and~II.
The result \eqref{eq:diffcharge} can be derived from requiring functional differentiability of the canonical Poincar\'e generator $\tilde G_P[\xi^\rho]$ \cite{Blagojevic:2010ir,ABEGJ:2011}.
Functional differentiability of the canonical Weyl generator $\tilde G_W[\Om]$ in general also leads to Weyl charges $Q_W[\Om]$, which, however, are trivial for cases~I and~II.

\section{Varying Weyl factor}

Case III has several similarities to case~II, but also some essential differences.
For the same reason as before we restrict the Weyl factor $\phi$ as in \eqref{eq:gCS30}, but with $f$ now being free rather than fixed. 
The gauge transformations preserving the boundary conditions still include the asymptotic diffeomorphisms \eqref{eq:gCS28}, while the allowed Weyl rescalings now include all $\Om$ of the form
\eq{
\Om =  \Om(x^+,\,x^-) + {\cal O}(y^2)
}{eq:gCS27}
with an arbitrary function $\Om(x^+,\,x^-)$. 
Relatedly, we allow variations of the Weyl factor $\de\phi$ with an arbitrary function $\de f(x^+,\,x^-)$. 
\eq{
\de \phi = \de f(x^+,\,x^-) + \ldots
}{eq:free2}

Let us start with stating the canonical result for the Weyl charge \cite{ABEGJ:2011}, which for case~III becomes non-trivial.
\eq{
\de\,Q_W[\Om] = -\frac{k}{\pi} \int\limits_0^{2\pi} \extd \varphi \, \de f \, \partial_\varphi \Om \, .
}{eq:free8}
Clearly, the charge \eqref{eq:free8} is not conserved for arbitrary functions $f$ and $\Om$. 
It is conserved if and only if $\partial_t  (f \partial_\varphi \Om )$
is a total $\varphi$-derivative.

Let us consider explicitly the case when the stress tensor is conserved, i.e., when the flatness condition \eqref{eq:an7.5} holds.
To preserve this form, we must also have $\Omega = \Om_+(x^+) + \Om_-(x^-)$. 
Fourier expanding these functions as
\begin{align}
f_\pm &= \frac{f_0}{2} + \frac{p_f}{2}(t \pm \varphi) + \sum_{n\neq 0} f_\pm^{(n)} e^{-in(t \pm \varphi)}\, , \label{eq:free10} \\
\Om_\pm &= \frac{\Om_0}{2} + \frac{p_\Om}{2}(t \pm \varphi) + \sum_{n\neq 0} \Om_\pm^{(n)} e^{-in(t \pm \varphi)}\, , \label{eq:free11}
\end{align}
one finds straightforwardly that the conservation of the Weyl charge is equivalent to requiring
\eq{
f_+^{(n)} \Om_-^{(n)} = f_-^{(n)} \Om_+^{(n)}\qquad \forall n \neq 0  \,.
}{eq:free12}
This means that the Weyl rescaling $\Omega$ preserves some functional relation between $f_+$ and $f_-$ of the form
\eq{
f_+^{(n)} = C_n f_-^{(n)}
}{eq:free13}
for some constants $C_n$. Particularly simple choices are
\eq{
 f_- = 0\, ,  \quad f_+ = 0\, ,  \quad\mbox{or} \quad f_+(x) = C\, f_-(x)\,.
}{eq:free14}
Any such choice leads to an infinite tower of conserved charges. 
For instance, the first choice in Eq.~\eqref{eq:free14} leads to the tower
\eq{
Q_W[\Om = -e^{in(t+\varphi)}] = 2i k\, n f_+^{(n)} \, .
}{eq:free15}
The generators $\aff_n = \tilde{G}_W[\Om = -e^{in(t+\varphi)}]$ obey a simple Dirac bracket algebra.
\begin{align}
& 
i\,\{\aff_n, \aff_m\}^\ast = 2k n\,\de_{n+m,0} \label{eq:gCS89} 
\end{align}
The asterisk denotes Dirac brackets defined in a specific partially reduced phase space \cite{ABEGJ:2011}.

Let us now turn to the Virasoro charges.
The asymptotic expansion of the dreibein $\bar e^i{}_\mu:=\exp{(-\phi)}\,e^i{}_\mu$ reads
\eq{
\bar e^{\, i}{}_\mu = \frac{1}{y}\, \de^i_\mu + \begin{pmatrix} 0&0&0\\\bar e_{(1)+}^{\, -} &0 &0 \\ 0&0&0 \end{pmatrix}
 +  y \begin{pmatrix} \bar e_{(2)+}^{\, +} &\bar e_{(2)-}^{\, +}&0\\\bar e_{(2)+}^{\, -} &\bar e_{(2)-}^{\, -} &0 \\ 0&0&0 \end{pmatrix} + \ldots
}{eq:lalapetz}
and similarly for $\de\bar e^i{}_\mu$.
The expansions for the dualized spin-connection $\om^i$ and the Lagrange multiplier 1-form $\la^i$ then follow straightforwardly from the equations of motion descending from the first order action \eqref{2.1}.
Putting these expressions into the variation \eqref{eq:diffcharge} produces
\begin{multline}
\de Q_P[\xi^\rho] = -\frac{k}{\pi}\,\int\limits_0^{2\pi} \extd \varphi \, \Big[ \xi^+ \, \de C_+  + \xi^-\, \de C_-  + \xi^y\, \de C_y \\
+\theta^{\hat y}[\xi^\rho]\,\de\om_{\hat y\varphi}\Big]\, , 
\end{multline}
with 
\begin{align}
&\de C_\pm =  \de\bar e_{(2)\pm}^\mp + \partial_\pm \partial_\varphi \de \phi - (\partial_\pm \phi) \partial_\varphi \de\phi + {\cal O}(y)\, , \\
&\de C_y = \frac{1}{y}\partial_\varphi \de\phi + {\cal O}(1)\, , 
\end{align}
\begin{align}
&\theta^{\hat y}[\xi^\rho]\,\de\om_{\hat y\varphi} = \frac12\,\big(\partial_+\eps^+-\partial_-\eps^-\big)\partial_t\de\phi + {\cal O}(y)\,. 
\end{align}
To reduce clutter we consider a Virasoro transformation with only $\eps^+$ nonzero. (The formulas are completely analogous
for the general case.) The diffeomorphism charges become
\begin{equation}
\de Q_P[\xi^\rho] = -\frac{k}{\pi} \int\limits_0^{2\pi} \extd \varphi \, \eps^+ \, \Big[ \de\bar e_{(2)+}^{\,-}  - (\partial_+ \phi) \partial_\varphi \de\phi \Big] \, ,
\end{equation}
where we dropped a total $\varphi$-derivative.
Note that the second term ($\sim \partial_+ \phi \partial_\varphi\de \phi$) is not integrable in general. However, considering a compensating 
Weyl rescaling as in case II we have
\begin{equation}
\de Q_W[\xi^\rho] = -\frac{k}{\pi} \int\limits_0^{2\pi} \extd \varphi \, \left[ -\de \phi \, \partial_\varphi(\eps^+\partial_+ \phi)\right] \, .  \label{eq:gCS94}
\end{equation}
This expression results from combining \eqref{eq:free8} and \eqref{eq:gCS29}. 
Adding these contributions, $\de Q[\xi^\rho] = \de Q_P[\xi^\rho] + \de Q_W[\xi^\rho]$, turns the terms bilinear in $\phi$ and $\de \phi$ into a $\varphi$-derivative. 
Taking into account also the terms proportional to $\eps^-$ we obtain the charges
\eq{
Q[\xi^\rho] = -\frac{k}{\pi}\, \int\limits_0^{2\pi} \extd \varphi \, \Big[\eps^+ \,  \bar e_{(2)+}^{\,-}  + \eps^- \,  \bar e_{(2)-}^{\,+}\Big]\, .
}{eq:gCS88}
The result \eqref{eq:gCS88} coincides with the one for cases~I and II. Thus, the canonical charges \eqref{eq:gCS88} are conserved.

\section{CFT interpretation}\label{se:5}

Based upon the results above we conjecture that CSG with boundary conditions \eqref{eq:bc1}-\eqref{eq:bc3}, \eqref{eq:gCS30} and $f=f(x^+)$ is dual to a CFT with enhanced symmetries.
These symmetries are generated by the Virasoro operators $L_n$, $\bar L_n$ as well as the generators $\aff_n$. The Virasoro generators are defined as a combination of diffeomorphisms \eqref{eq:gCS28}  accompanied by
compensating Weyl rescalings \eqref{eq:gCS29}: $L_n=\tilde G_P[\eps^+=e^{inx^+}]+\tilde G_W[\Om=\Om(\eps^+)]$ and $\bar L_n=\tilde G_P[\eps^-=-e^{-inx^-}]+\tilde G_W[\Om=\Om(\eps^-)]$. 
Therefore the non-zero commutators are
\begin{subequations}
\label{eq:algebra}
\begin{align}
 [L_n,\,L_m] &= (n-m)\,L_{n+m} + \frac{c_R}{12}\,(n^3-n)\,\de_{n+m,0} \, , \label{eq:CFT1}\\
 [\bar L_n,\,\bar L_m] &= (n-m)\,\bar L_{n+m} + \frac{c_L}{12}\,(n^3-n)\,\de_{n+m,0}\, , \\
 [\aff_n,\,\aff_m] &= c_0\,n\,\de_{n+m,0} \, .
\end{align}
\end{subequations}
The values of the central charges are determined by the Chern--Simons level $k$ from the action \eqref{eq:gCS10}:
\eq{
 c_R = - c_L = 6\, c_0 = 12\,k
}{eq:gCS86}
Of course, the value of $c_0$ is defined only with respect to a given normalization of the generators $\aff_n$.
The commutator $[L_n,\,\aff_m]$ vanishes due to the peculiar way the Virasoro generators $L_n$, $\bar L_n$ arise.
By construction they act trivially on the 3-dimensional Weyl factor.

We define now Sugawara-shifted generators $\vir_n$ that generate only holomorphic diffeomorphisms ($::$ denotes normal ordering):
\eq{
L_n\to\vir_n = L_n + \frac{1}{4k}\,\sum_{m\in\mathbb{Z}} :\aff_m \aff_{n-m}:
}{eq:gCS93}
To show that the generators $\vir_n$ produce only diffeomorphisms it is sufficient to check that the compensating Weyl charge \eqref{eq:gCS94} by virtue of \eqref{eq:free15} can be written as
\begin{multline}
\de Q_W = \frac{k}{\pi}\,\int\limits_0^{2\pi}\extd\varphi\, \de f \partial_\varphi ( e^{inx^+}\partial_+ f) \\
= k\,\de \Big(\sum_{m\in\mathbb{Z}} m(n-m) f_+^{(m)}f_+^{(n-m)}\Big) \Rightarrow \\
\tilde G_W = - \frac{1}{4k}\,\sum_{m\in\mathbb{Z}} :\aff_m \aff_{n-m}:
\label{eq:gCS95}
\end{multline}
In the last equality we have converted classical expressions into quantum operators, with corresponding ordering ambiguities.
We have fixed the latter by requiring normal ordering.
Since $L_n$ is a sum of a diffeomorphism and a holomorphic Weyl rescaling with Weyl charge \eqref{eq:gCS95}, the shifted Virasoro operators $\vir_n$ in \eqref{eq:gCS93} generate by construction solely diffeomorphisms.
The definition \eqref{eq:gCS93} together with the old algebra \eqref{eq:algebra} establish a new algebra that contains an affine $\hat u(1)_R$ generated by $\aff_n$.
\begin{subequations}
\label{eq:algebranew}
\begin{align}
 [\vir_n,\,\vir_m] &= (n-m)\,\vir_{n+m} + \big(k+\frac{1}{12}\big)\,(n^3-n)\,\de_{n+m,0} \label{eq:CFT1new}\\
 [\bar L_n,\,\bar L_m] &= (n-m)\,\bar L_{n+m} - k\,(n^3-n)\,\de_{n+m,0}\\
 [\aff_n,\,\aff_m] &= 2k\,n\,\de_{n+m,0} \\
 [\vir_n,\,\aff_m] &= -m\,\aff_{n+m} 
\end{align}
\end{subequations}
Note in particular that the last commutator is now non-vanishing and shows that $\aff_n$ behaves as an operator with the appropriate conformal weights $(1,0)$.
As compared to cases I and II, in case III the holomorphic Weyl factor constitutes an additional free chiral boson in the theory.
The shift of $c_R\to c_R+1$ (and no shift of $c_L$) is precisely what one would expect from a free chiral boson.
Note that there is no corresponding shift of the left central charge or the left Virasoro generators, since charge conservation demands that we accompany a diffeomorphism generated by $\xi[\varepsilon^-=-e^{inx^-}]$ with a compensating  anti-holomorphic Weyl rescaling \eqref{eq:gCS29} with $\varepsilon^-=-e^{inx^-}$. 

It is worthwhile pointing out the peculiar way in which the chiral boson arises on the gravity side.
In the bulk there is no scalar field, but only the Weyl factor $\phi$ in \eqref{eq:bc1}.
The bulk equations of motion \eqref{eq:gCS1} do not restrict $\phi$ at all!
The whole dynamics of $\phi$ emerges through consistency conditions imposed at the boundary.
If the stress-energy tensor is postulated to be conserved then the flatness condition \eqref{eq:an7.5} must hold, which is equivalent to demanding that $\phi$ obeys the massless Klein-Gordon equation at $\partial M$.
If additionally the Weyl charges are required to be conserved then $\phi$ is restricted further, as we have shown above.
Thus, the whole dynamics of the scalar field arises solely through boundary and consistency conditions, and not through an interplay between bulk and boundary dynamics.
It will be interesting to relax some of these requirements and study consequences for the dual CFT \cite{ABEGJ:2011}.
For instance, considering a curved boundary metric with Ricci scalar ${\cal R}$ we expect the condition \eqref{eq:an7.5} to be replaced by the Liouville equation $\nabla^2 f \propto {\cal R}$.
Another interesting generalization is to allow for Weyl factors $\phi$ that contain a term $\exponent\,\ln{y}$ (see footnote~[18]).

For the second choice in \eqref{eq:free14} essentially the same discussion applies, with suitable changes $L\leftrightarrow R$.
Other choices of the constants $C_n$ in \eqref{eq:free13} lead to correspondingly different dual CFTs.
One particular class of choices exhibits an interesting feature.
Assuming
\eq{
C_n C_{-n} = |C_n|^2 = 1 \qquad \forall n
}{eq:restriction}
we find that all Weyl charges vanish and the generators $\aff_n$ commute with each other. 
A more comprehensive discussion of these and more general choices, as well as their consequences, will be presented elsewhere \cite{ABEGJ:2011}.

Some of the cases~II (and case~I) can be interpreted as constant Weyl charge superselection sectors of a specific case~III CFT.
For instance, case~II with holomorphic Weyl factor $\Om(x^+)$ is recovered from the CFT defined by \eqref{eq:algebra}-\eqref{eq:gCS86} by choosing $\Om$ correspondingly and by setting all Weyl charges to zero.
If $\Om$ is constant then case~I is recovered.
However, not all cases~II can be recovered in this way, because of the restriction \eqref{eq:free13}. 

The CFTs discussed here cannot be unitary since one of the central charges always is negative (except for some non-integer $k$), e.g.~$c_L<0$.
Redefining $\bar L_n\to-\bar L_{-n}$ makes the central charge positive, but the corresponding CFT still is not unitary, since there are states with negative weights, as evident from the spectrum of Weyl gravitons, see \eqref{eq:JJ}. 
It remains a challenge to construct some unitary dual CFT for CSG.

Finally, it would be of interest to generalize our results, where applicable, to 4-dimensional conformal gravity, see \cite{Lu:2011zk,Maldacena:2011mk} and references therein.

\bigskip

\acknowledgments

We thank Radoslav Rashkov and Thomas Zojer for discussions.
HA thanks Hessamaddin Arfaei for his support and encouragement as well as ITP members and secretaries in Vienna for their hospitality.

HA is supported by the Ministry of Science, Research and Technology in Iran, and during the final stages also by the START project Y435-N16 of the Austrian Science Fund (FWF).
BC is supported by the Serbian Science Foundation, Grant No.~171031.
SE, DG and NJ are supported by the FWF projects Y435-N16 and P21927-N16. 
BC acknowledges travel support from the FWF project Y435-N16.


\end{document}